\documentclass[useAMS,usenatbib]{mn2e}
\usepackage{graphicx}
\usepackage{xspace}

\newcommand{\araa}{ARA\&A}
\newcommand{\apj}{ApJ}
\newcommand{\apjl}{ApJ}
\newcommand{\aap}{A\&A}
\newcommand{\mnras}{MNRAS}

\newcommand{\bb}[1]{\ifmmode \mbox{\boldmath $ #1$} \else  \mbox{\boldmath $#1$} \fi}

\newcommand{\U}[1]{\ensuremath{\mathrm{~#1}}}

\newcommand{\yr}{\U{yr}}
\newcommand{\Myr}{\U{Myr}}
\newcommand{\Gyr}{\U{Gyr}}
\newcommand{\pc}{\U{pc}}
\newcommand{\kpc}{\U{kpc}}

\newcommand{\msun}{\U{M}_{\odot}}
\newcommand{\Msun}{\msun}
\newcommand{\kms}{\U{km\ s^{-1}}}

\newcommand{\rv}{r_{\rm v}}

\newcommand{\nbody}{\texttt{NBODY6}\xspace}
\newcommand{\nbodytt}{\texttt{NBODY6tt}\xspace}

\newcommand{\fig}[2][]{Figure#1~\ref{fig:#2}}

\renewcommand{\fig}[2][]{Fig#1.~\ref{fig:#2}}

\title{The role of galaxy mergers on the evolution of star clusters}

\author[Renaud \& Gieles]{Florent~Renaud$^1$\thanks{florent.renaud@cea.fr}, Mark~Gieles$^{2,3}$\\
$^1$ Laboratoire AIM Paris-Saclay, CEA/IRFU/SAp, Universit\'e Paris Diderot, F-91191 Gif-sur-Yvette Cedex, France\\
$^2$ Institute of Astronomy, University of Cambridge, Madingley Road, Cambridge, CB3 0HA, UK\\
$^3$ Department of Physics, University of Surrey, Guildford GU2 7XH, UK\\
}
\date{Accepted 2013 January 29.  Received 2013 January 28; in original form 2013 January 9}
\begin{document}
\maketitle

\defcitealias{Renaud2011}{RGB11}

\begin{abstract}
Interacting galaxies favor the formation of star clusters but are also suspected to affect their evolution through an intense and rapidly varying tidal field. Treating this complex behaviour remains out-of-reach of (semi-)analytical studies. By computing the tidal field from galactic models and including it into star-by-star $N$-body simulations of star clusters, we monitor the structure and mass evolution of a population of clusters in a galaxy major merger, taking the Antennae galaxies (NGC~4038/39) as a prototype. On the long timescale ($\sim 10^9 \yr$), the merger only indirectly affects the evolution of clusters by modifying their orbits in or around the galaxies: the mass-loss of clusters in the merger remnant is faster, while clusters ejected in the tidal debris survive much longer, compared to in an isolated galaxy. The tidal perturbations of the galactic collisions themselves are too short lived and not strong enough to significantly influence the structure and dissolution of realistically dense/massive star clusters.
\end{abstract}

\section{Introduction}

For 40 years now, since the pioneer work of \citet{Toomre1972}, many studies focussed on the role of mergers on galaxy evolution \citep[among many others, see][]{Elmegreen1993, Bournaud2008, Lotz2008}. Hydrodynamical simulations explored the process of star formation \citep{Kravtsov2005}, with the goal of interpreting the observational data of (Ultra-) Luminous Infra-Red Galaxies, and formation of star clusters often detected shortly after a major merger event \citep{Ashman1992, Brodie2006}. Numerical simulations confirmed indeed that galaxy-galaxy interactions could significantly increase the star \emph{cluster} formation rate \citep[e.g.][]{diMatteo2007, Karl2010, Teyssier2010}. However, at the stage of the conversion of gas into stellar material, a star-by-star exploration in a full galactic context, but also the subsequent dynamical evolution, remain technically out-of-reach.

However, the fate of these clusters after their formation remains uncertain. This is because an evaporating/dissolving cluster in a galaxy beyond the local group leaves no clear observational signature for us to find. Numerical studies are challenged because they must account for a wide range of scales, from the internal dynamics of clusters (i.e. binary stars and close encounters) to the scale of galactic interactions. This issue has been circumvented by simplifying either one or the other side of the problem. On the one hand, an idealized galaxy with an analytical description of the tides has been intensively used to improve our understanding of the dynamical evolution of non-isolated clusters \citep[e.g.][]{Vesperini1997, Baumgardt2003, Penarrubia2009, Kuepper2010a, Berentzen2012, Webb2013}. On the other, a semi-analytical description of star clusters in a cosmological context allowed to probe the effect of complex tides on stellar populations \citep[e.g.][]{Prieto2008, Kruijssen2011}.

Recently, \citet*[][hereafter RGB11]{Renaud2011} merged the two approaches by integrating any tidal effect into $N$-body simulations of star clusters, allowing for the exploration of the evolution of star clusters within complex, time-dependent galactic potentials, like those of mergers. The goal of this Lettre is to apply this method to a large number of clusters in rapidly-varying tidal fields, and to derive the differences in their evolution and dissolution rate between isolated galaxies and mergers. Our approach focusses first on a few cases to better understand the physical phenomenon of dissolution itself, and then on an entire population of stars clusters. This work does not intent to reproduce realistic clusters, but rather to shed light on the physical effect of complex time-evolving tides, leaving aside, for the time being, other aspects like mass segregation and stellar evolution.

\section{Methodology}
\label{sec:methodology}

\subsection{Numerical method}
\label{sec:numericalmethod}

The evolution of our cluster models is computed with the code\footnote{http://irfu.cea.fr/Pisp/florent.renaud/nbody6tt.php} \nbodytt, widely based on \nbody \citep{Aarseth2003,Nitadori2012} and presented in \citetalias{Renaud2011}, and briefly summarised here. First, a simulation of a galaxy is performed: each particle represents a possible star cluster. (This step can be done using any type of gravitational code, e.g. tree-code.) One particle is followed along its orbit and the associated tidal tensors are extracted \citep[see][]{Renaud2008}, providing a complete space and time description of the tides for this cluster candidate. Next, in a second simulation, \nbodytt creates a star-by-star $N$-body model of a star cluster, reads the tensors, interpolates them in time, computes the tidal forces at the positions of the $N$ stars of the cluster and adds them to the internal gravitational force due to the $N-1$ other stars. A significant fraction of the computation is speed-up thanks to the use of Graphics Processing Units (GPUs). Details and tests can be found in \citetalias{Renaud2011}. In this Lettre, we repeat this exercise by considering various orbits of clusters, i.e. various tidal fields, in a merger and in an isolated galaxy. To focus on the tidal effect and on the dissolution of the cluster population, we do not include any prescription on the formation of our clusters. In total, 724 $N$-body simulations have been run, on the GPU cluster \emph{Curie} hosted at the \emph{Tr\`es Grand Centre de Calcul} (TGCC).

\subsection{Galaxies, orbits and clusters}
\label{sec:descriptionoftheorbits}

We computed the tidal tensors using a tree-code along several orbits in a purely gravitational model of NGC~4038. (The method and simulation details are given in \citealt{Renaud2008}.) This galaxy has been modeled both in isolation and with its equal-mass companion (NGC~4039), the pair reproducing the morphology and kinematics of the Antennae galaxies. This allows a direct comparison between the isolated galaxy and the major merger. 

As the galactic collisions proceed, the tidal history gets punctuated by ``events'' corresponding to the impact of the two galaxies. These events however are independent of the intrinsic evolution of the clusters, itself paced by two-body relaxation \citep{Henon1961, Gieles2010c}. To avoid any artificial synchronisation between the galaxy and the cluster, we start the integration of the cluster models at different times with respect to the collisions, i.e. at different positions along the orbits. This way, our sample comprises an entire population of young, intermediate and old clusters. However, all the simulations are stopped $5 \Gyr$ after the first galactic pericenter passage: for longer timelapses, cosmological aspects stepping in the galaxy evolution should be taken into account, which is out of the scope of the present study. With 32 orbits and five ``birth'' epochs along them, we can monitor histories representative of all cases observed in a merger: clusters confined to the galaxy nucleus, or ejected in the close vicinity of the central region, in tidal bridges or tails, falling back and orbiting the remnant, or being definitively ejected in the intergalactic medium, and this at different stages of their evolution. Half of the 32 orbits have been selected to be confined to the galactic disc (before the interaction occurs) while the others reach positions in the stellar halo above and below the plane of the disc, on very inclined orbits. The maximum apocenter corresponds to the radius of the stellar disc ($20 \kpc$) such that our survey represents a full range of orbital energies (see Section~\ref{sec:clusterpopulation}).

For simplicity, in each cluster, all stars have been given the same mass ($1 \msun$), to avoid mass segregation and stellar evolution effects and to focus on the tides. The initial positions and velocities of the stars were based on a virialised \citet{Plummer1911} model. On each orbit and for each birth epoch, we set up clusters with different masses and densities obtained either by changing the initial number $N_0$ of stars per cluster from 4000 to 32000 for a virial radius $\rv = 1\pc$, or by changing $\rv$ (down to $0.5 \pc$) for a given $N_0$. The half-mass densities vary from $10^3$ to $10^4 \Msun\pc^{-3}$, which makes our models comparable in mass and density to Westerlund~1, NGC~3603 or even the Arches for the densest ones \citep{Portegies2010}. However, because of early evolution, the densities get lower (down to two orders of magnitude in the extreme case) at the time of the galactic interaction.

\section{Results}
\label{sec:results}

By looking at the mass evolution curves of all individual clusters in our survey, we found shows that every case is unique. However, it is possible to extract families of clusters sharing a similar tidal history.

\subsection{Individual cases}
\label{sec:individualcases}

\begin{figure}
\includegraphics{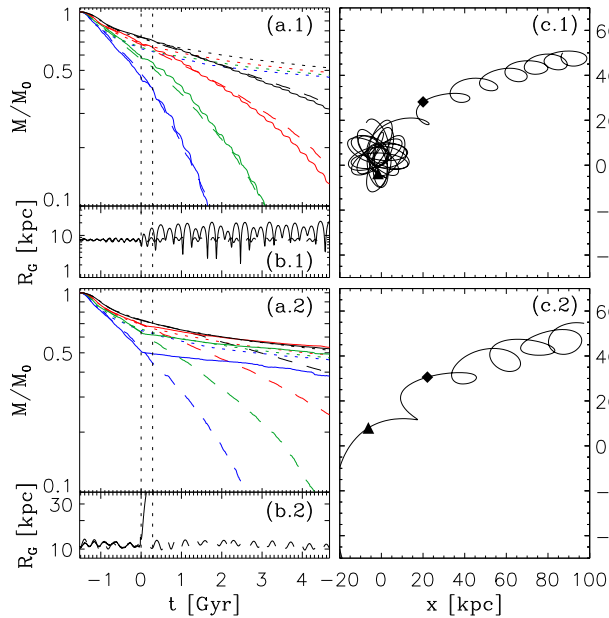}
\caption{Evolution of clusters along two orbits arbitrarily selected. Panels (a): normalized mass of the clusters, without tides (dotted curves), in the isolated galaxy (dashed) and in the merger (solid). The colour indicates the initial mass of the cluster (4000, 8000, 16000 and 32000 stars for blue, green, red and black respectively, all with an initial virial radius of $1 \pc$). Vertical dotted lines mark the galactic pericentre passages; the second one corresponds to the final coalescence. Panels (b): Galactocentric distance. Panels (c): Orbit of the clusters in the orbital plane of the merger. (The clusters start from the top-right corner. A diamond and a triangle indicate the position of the clusters at the times of the two galactic collisions. See \citealt{Renaud2009} for details on the galactic simulation.)}
\label{fig:examples}
\end{figure}

First, in \fig{examples}, we show two orbits from our survey, in order to analyse typical situations that may occur at galactic scale and their implications for the clusters. Only the simulations with a birth epoch of $\sim 1.5 \Gyr$ before the first pericentre passage are shown here, for the sake of clarity, but the other runs give comparable results.

On the top row case (panels abc.1), the orbital eccentricity is rapidly increased from $\sim 0.05$ to $\sim 0.30$ by the galactic collision. However the trajectory remains closed and fairly regular. The resulting mass evolution shows the typical ``staircase'' decline, the mass-loss becoming more rapid at small galactocentric radii because of strong gravitational tides and a significant centrifugal force. The amplitude of this periodically accelerated mass-loss increases with low-mass clusters. This is also found for eccentric orbits (e.g. \citealt{Baumgardt2003}, \citetalias{Renaud2011}). Despite these short-period ($\sim 10^8 \yr$) changes induced by the merger, the secular evolution of this cluster does not strongly differ from that it has in an isolated disc (dashed lines on \fig{examples}). In that sense, the merger does not affect the mass-loss rate of clusters \emph{on this particular orbit}. In Section~\ref{sec:clusterpopulation}, we show that this is not a representative example, in fact, because the mass-loss rate of clusters that remain bound to the merger remnant increases on average.

On the contrary, the orbit of our second example (panels abc.2) is much more altered by the galactic collision. At the first galactic pericentre passage ($t=0$), the orbit follows the creation and expansion of a galactic tidal tail and does not fall back into the nucleus. Being in a shallow potential, the clusters on this trajectory experience a very weak tidal field, and thus a slow mass-loss (the mass-loss rate for $t > 0.3 \Gyr$ is comparable to that of an isolated cluster, i.e. most escape happens because of close encounters in the core rather than a relaxation driven evaporation over the tidal boundary), whereas in the isolated disc, the orbit remains bound and the clusters dissolve in a few $\times 10^9 \yr$. In this case, the merger protects the clusters by sending them to a zone of weaker tides.

\subsection{Cluster population}
\label{sec:clusterpopulation}

To generalise these results, we now consider all our clusters, but tell apart two sub-populations: the clusters always staying within $20 \kpc$ from the center of the galaxy (i.e. about the radius of the isolated disc) and the others, ejected to the tidal debris. About $30\%$ of the orbits fall in this second category, which corresponds to the stellar mass fraction lost by the progenitor galaxies of the Antennae during the interaction. This demonstrates that our sample of orbits is statistically representative. (For the isolated disc, this distinction is not made since all the clusters remain bound to their host galaxy.) 

\subsubsection{Mass-loss}
\label{sec:massloss}

\begin{figure}
\includegraphics{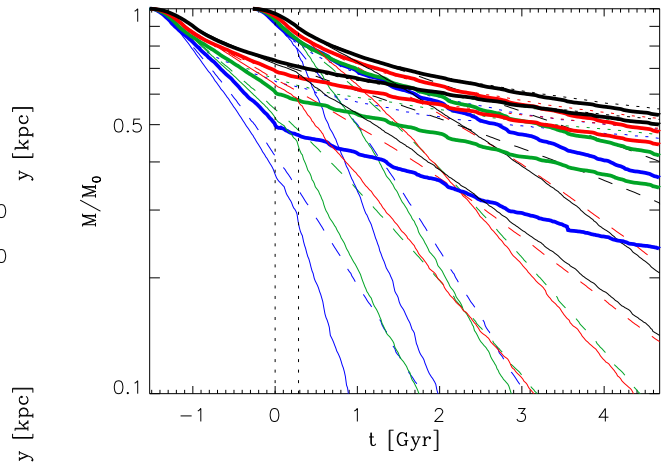}
\caption{Evolution of the normalized mass averaged by telling apart the clusters ejected to large galactocentric radii by the galaxy-galaxy collision (solid thick line) from those remaining close to the center (solid thin line). The dashed lines represent all the clusters in the isolated disc and the dotted lines denotes the mass-loss without tides. As in \fig{examples}, the colour indicates the initial mass of the cluster.}
\label{fig:average_groups}
\end{figure}

\fig{average_groups} shows the average mass for the two groups and for two birth epochs. The ``staircase'' behaviour is smoothed out because all the orbits in our sample which might exhibit it are not synchronised. As expected from the example of the previous Section, the ejected clusters (thick solid line) live much longer as the tidal field they experience is weaker than in the merger remnant and the isolated galaxy. The knees in their mass evolution, marking a change of tidal regime, occur at first pericentre passage, when the tidal tails are created. However, at that time, the non-ejected clusters only experience the effect of the cumulated potentials of the two galaxies (before they separate and fall-back onto each other), which is too short-lived to significantly affect the mass-loss (see Section~\ref{sec:collisions}). However, at the final coalescence ($t=0.3 \Gyr$) the curve of the mass evolution yields a sharp knee. Contrarily to the particular case presented in the top row of \fig{examples}, on average the mass of the non-ejected clusters decreases much more rapidly in the merger than in the isolated disc. Clusters in the central region of a merger remnant are very likely to dissolve faster (within a few Gyr, for the masses considered here) than in an isolated galaxy, partly because of an increased average orbital eccentricity from $0.09 \pm 0.03$ before the first collision, to $0.39 \pm 0.12$ after the merger. 

\begin{figure}
\includegraphics{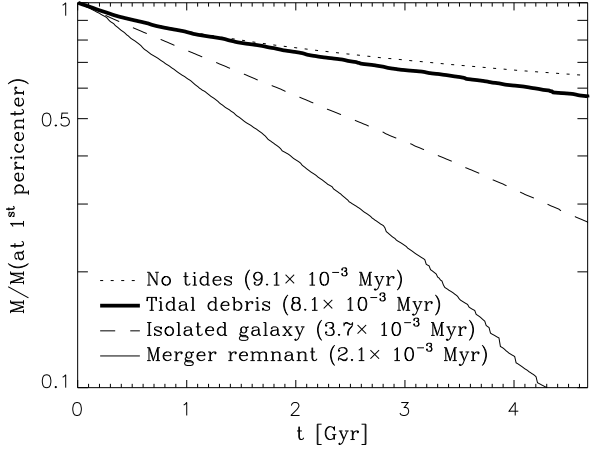}
\caption{Dissolution of the entire population of clusters (all densities, all birth epochs and all orbits) for our four families of tidal histories since the first galactic collision. The mass is normalized to its value at the first pericenter passage. The number in parenthesis is the characteristic timescale $\tau$ of the exponential decay of the mass ($M \propto e^{-t/\tau}$).}
\label{fig:population}
\end{figure}

Finally, \fig{population} compares the evolution of the mass of the entire population of a galaxy, independently of the cluster density, mass or birth epoch, but still telling apart the ejected clusters (tidal debris) from those staying in the remnant. The pre-merger history ($t<0$) is not shown here because it encompasses the artificial creation of clusters at arbitrary times and thus, is not physically relevant for our comparisons. In all cases, the average mass decays as $M \propto e^{-t/\tau}$ up to a very late stage of the evolution. The dissolution rate $-1/\tau$ of the population in the tidal debris is comparable (yet a bit higher) to that of the tide-free clusters and $\sim 4$ times lower than in the merger remnant. The isolated galaxy exhibits an intermediate dissolution rate, twice lower than that of the merger remnant. From the results in \fig{population}, we thus conclude that the average mass-loss rate increases for clusters that remain bound to the merger remnant. For some clusters the mass-loss stays comparable (e.g. the example in the top panel of \fig{examples}), but it increases for a majority of clusters.

\subsubsection{Structural evolution}
\label{sec:structuralevolution}

\begin{figure}
\includegraphics{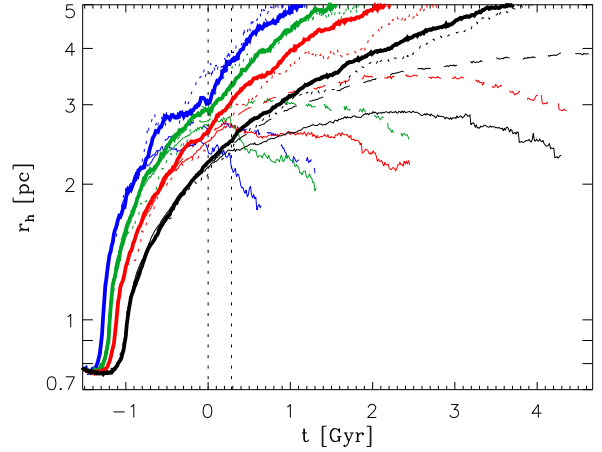}
\caption{Evolution of the half-mass radius of our clusters. The curves are stopped when the corresponding average normalized mass is less than 0.15, making the half-mass radius very Poisson-noisy. Colors and linestyles are as in \fig{average_groups}.}
\label{fig:rh}
\end{figure}

In this Section, we focus on the structural changes along the evolution of our models. \citet{Gieles2011b} explained that, after a relaxation-driven expansion phase, clusters in a constant tidal field experience a contraction phase. In \fig{rh}, we show that clusters in time-varying fields also experience a comparable two-phase evolution. The transition from expansion to contraction occurs half-way the lifetime of the cluster, as predicted by \citet{Gieles2011b} for steady tides. This demonstrates that the variations of the tidal field ($\sim 10^8 \yr$), i.e. rapid compared to the lifetime of the cluster ($\sim 10^9 \yr$), do not modify qualitatively the behaviour from what is expected in constant tides.

\begin{figure}
\includegraphics{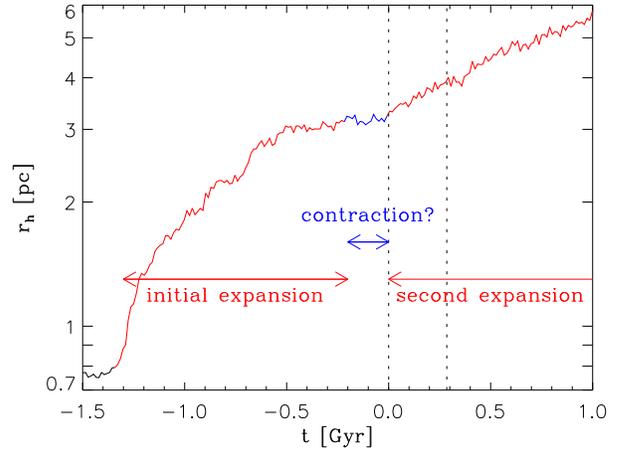}
\caption{Evolution of the half-mass radius of a cluster sent in the tidal debris (one cluster of the group shown by the blue solid thick line in \fig{rh}). The new tidal regime starting at $t=0$ allows for a resumption of the expansion phase and leads to an extended cluster.}
\label{fig:double_expansion}
\end{figure}

As noted above, the clusters remaining in the central region of the merger (thin solid lines) experience stronger tides than in the isolated galaxy (dashed lines) which makes their contraction more important. The clusters ejected into the tidal debris (thick solid lines) follow a quasi tide-free evolution and thus miss the tidally-driven contraction phase. In \fig{double_expansion}, we plot the evolution of the half-mass radius of one cluster set on the orbit shown in \fig{examples} (bottom panels). At the time of the first collision, the expansion phase of this cluster is over and the contraction has just begun. However, the sudden change of tidal regime due to the ejection to the tidal tails stops the contraction and allows for a much longer expansion.

\subsection{Role of collisions}
\label{sec:collisions}

In the previous Section, we showed the indirect changes implied by the merger on its population through orbital mixing, i.e. a long-term effect. However, the tides are expected to have their maximum influence at the time of the galactic pericenter passages, when the gravitational potentials of the still undisturbed galaxies most overlap. At the collision ($t=0$), the velocity difference between the two galaxies is $\sim 200 \kms$, therefore a significant fraction of their masses overlaps for a comparable or shorter timelapse than the crossing times of our clusters ($\sim 10\--100 \Myr$). Furthermore, it should be noted that the geometry of the mergers does not necessarily account for a global strengthening of the tides: one galaxy can locally compensates for the tides of the other by flattening the gravitational potential\footnote{In the idealised case of two point-mass galaxies, the effective potential has a saddle point where the tidal forces balance the gravitation of the cluster. For more complex galaxies, this situation may occur over larger volumes, depending on the shape of the potential and the geometry of the encounter.}. In short, during the collision, the tidal field could be locally weaker than expected, but above all, the combination of the effects of the two galaxies is short-lived compared to the crossing time of their clusters.

In our survey, the more fragile clusters loose only $\sim 2\%$ of their current mass during the first galactic collision. For instance, the clusters like that of \fig{double_expansion} have an initial half-mass density of $1000 \msun\pc^{-3}$, which becomes $\sim 10 \msun\pc^{-3}$ at the time of the collision. Even on such low density objects, the direct effect of the collision is very mild. It is visible in the mass-loss curves (see the blue curves in \fig{examples}.a.1 and \fig{population} around $t=0$) but is negligible with respect to the classical rapid mass-loss observed when the cluster reaches the pericenter of its orbit (i.e. the steep part of the ``staircase'' shape mentioned above). Although the orbital changes occurring in the merger remnant makes a similar analysis much more difficult at the second collision ($t \approx 0.3 \Gyr$), we expect qualitatively this behaviour to repeat itself. \emph{In fine}, the galactic collisions themselves have no direct impact on the mass-loss of clusters.

\section{Discussion and conclusion}
\label{sec:discussion}

We have studied the tidal effect of a galaxy major merger on its population of star clusters. Our main findings are:
\begin{itemize}
\item the mass-loss rate of the clusters that remain bound to the merger remnant is higher than before the final coalescence phase. The clusters populating the tidal debris survive much longer, similarly to tide-free cases.
\item the two-phases evolution of expansion and contraction of a cluster also exists in complex, time-varying tidal fields. The structural evolution does not strongly deviate from that of a constant tides case.
\item at the time of the galactic collisions, the tides are too weak (although they reach their maximum intensity) and too short-lived ($< 10^{7-8} \yr$) to have a significant influence on the clusters. However, by modifying the orbit of the clusters, they indirectly affect their mass-loss over long timescales.
\end{itemize}

By focussing on tidal effects, we have not considered stellar evolution, which influences the evolution of the mass and the structure of clusters. Stellar evolution causes clusters to expand faster in the first few $100 \Myr$, making clusters less dense and possibly more vulnerable to tidal forces. However, because in this study we have considered clusters with different physical ages, some of which already reached the tidal density before the merger, we do not expect that our overall conclusions will be affected by the inclusion of stellar evolution.

Furthermore, we have not taken the formation of the clusters into account. High resolution hydrodynamical simulations of the same galaxy model \citep{Teyssier2010} reveal that the star formation activity increases by a factor up to $\sim 20$ shortly after the pericentre passages when the turbulent interstellar medium is properly resolved. However, galaxy-scale simulations still treat star formation through subgrid recipes and most of them lack the resolution to properly describe the initial state of clusters. For theses reasons, our approach decouples the dissolution of the clusters from their formation, considering separately the artifacts from both sides. Hence, our results do \emph{not} represent the age or mass functions of clusters, but their mass-loss rate. Studying the formation sites of the objects presented here ($\sim 10^{3-4} \Msun$ clusters) in a fully consistent galactic context by resolving star forming cores and the stellar feedback down to $\sim 1 \pc$ is our next step (Renaud et al., in prep).

\section*{Acknowledgments}
We thank Morgan Fouesneau and Christian Boily for fruitful discussions, and the referee for a constructive report. This work made use of the resources of the TGCC under the allocations \emph{Grand~Challenge}~0002 and GENCI~6835, both run on the hybrid nodes of supercomputer Curie. FR acknowledges support from the EC through the grant ERC-StG-257720, and MG acknowledges the Royal Society for financial support.

\bibliographystyle{mn2e}

\end{document}